\documentclass[a4paper,11pt]{article}
\usepackage{jheppub} 

\def\beq#1\eeq{\begin{align}#1\end{align}}

\title{$\mathcal{N}=2$ SCFT with minimal flavor central charge}

\author[a,b]{Dan Xie}
\affiliation[a]{Center of Mathematical Sciences and Applications, Harvard University, Cambridge, 02138, USA}
\affiliation[b]{Jefferson Physical Laboratory, Harvard University, Cambridge, MA 02138, USA}

\abstract{We list 4d interacting $\mathcal{N}=2$ SCFT with minimal flavor central charge from the theory space constructed using 6d $(2,0)$ theory. For $ADE$ and $C_N$ flavor groups,
our theories saturate the bound found using bootstrap method, but other cases have  higher values. We find interesting rank one SCFTs with  $B_3, G_2, F_4, C_4\times U(1), C_1\times U(1)$ flavor symmetry. Many physical properties of these 
theories are also studied.}

\begin{document} 
\maketitle
\flushbottom

\section{Introduction}
Conformal central charges $(a,c)$ and flavor central charge $k_G$ are basic invariants for a four dimensional $\mathcal{N}=2$ SCFT. 
Using conformal bootstrap method, we have following constraints on those central charges:
\begin{itemize}
\item The ratio $a/c$ satisfies following constraint \cite{Hofman:2008ar}:  ${1\over2}\leq a/c \leq {5\over4}$, and the upper bound is saturated by free 
vector multiplets and the lower bound by free hypermultiplets. 
\item The central charge $c \geq {11\over 30}$ for interacting theory \cite{Liendo:2015ofa}, and this value is saturated by simplest Argyres Douglas theory \cite{Argyres:1995jj} (we call it $H_0$ theory to indicate that it has trivial flavor symmetry).  
\item The constraint on flavor central charge is more interesting \cite{Beem:2013sza}, and the bound is recovered in table. \ref{bound} (for interacting theory) \footnote{Our convention of flavor central charge is half of that used in \cite{Beem:2013sza}.}. 
There is also an interesting bound involving $c$ and $k_G$ \cite{ Lemos:2015orc}.
\end{itemize}

\begin{table}[h]
\begin{center}
\begin{tabular}{|c|c|c|c|c|c|}
\hline
$A_{N-1}$&$k\geq {N\over2}$ & $D_N$ & $k\geq N-2$ &$E_6$ & $k\geq 3$   \\ \hline
$E_7$ & $k\geq 4$ &$E_8$ & $k\geq 6$ &$\bf{B_N}$ & $k\geq N-{3\over2}$ \\ \hline
$C_N$ & ${N\over2}+1$ & $\bf{G_2}$ & $k\geq {5\over3}$ & $\bf{F_4}$ & $k\geq {5\over2}$ \\ \hline
\end{tabular}
\end{center}
\caption{The bound on flavor central charge, which comes from considering unitarity condition involving the operator $\hat{B}_2$. There are some exceptions for lower rank classical group case: the constraint is $A_{N-1} (N\geq 3),~ B_N(N\geq 4),~C_N (N\geq 3),~D_N(N\geq 4)$. The other bounds are: $A_1 (k_G\geq {4\over 3}),~D_2 /D_3 (k_G\geq 2),~B_2/C_2/B_3 (k_G\geq 2).$ }
\label{bound}
\end{table}

It is interesting to identify those theories saturating the flavor central charge bound. 
It is already known that the rank one AD theories of type $(A_1, A_2, D_4, E_6,E_7,E_8)$ \cite{Argyres:1995xn,Minahan:1996cj}  saturate
these bounds. The purpose of this paper is to identify the remaining cases. Our strategy is to scan the theory space constructed in \cite{Xie:2012hs,Wang:2015mra}, and identify 
the theory with minimal flavor central charge. Our findings are
\begin{itemize}
\item Our minimal theory saturates the flavor central charge bound for $G=ADE$ and $G=C_N$.
\item For $G=B_N,G_2, F_4$, our minimal theory does not saturate the bound in table. \ref{bound}. Our bound is $k_{B_N}\geq  N-1, ~k_{G_2}\geq 2,~ k_{F_4}\geq 3$.
\end{itemize} 
The conformal central charge $(a,c)$ and flavor central charge $k_G$, Coulomb branch spectrum for  these minimal theories are listed in table. \ref{data}. 
We also list whether there is an extra $U(1)$ flavor symmetry or not.

\begin{table}[h]
\begin{center}
\begin{tabular}{|c|c|c|c|c|c|}
\hline
Flavor group $G$& $k_G$&Coulomb branch &$a$&$c$ & Extra $U(1)$\\ \hline
$A_{N-1}$(N~even)& ${N\over2}$&$({N\over2}, {N-2\over2},\ldots, 2)$ & ${7 N^2-20\over 96}$ & $\frac{1}{12} \left(N^2-2\right)$ &Yes    \\ \hline
$A_{N-1}$(N~odd)& ${N\over2}$&$({N\over2}, {N-2\over2},\ldots, {3\over 2})$ & ${7 (N^2-1)\over 96}$ & $\frac{1}{12} (N-1) (N+1)$ &No    \\ \hline

$B_{N}$(N~even)& $N-1$&$(N-1,N-3 \ldots, 3)$ & ${7 N^2-5 N-10 \over 48}$ & $\frac{1}{12} \left(2 N^2-N-2\right)$ &Yes    \\ \hline
$B_{N}$(N~odd)& $N-1$&$(N-1,N-3 \ldots, 2)$ & ${ 7 N^2 - 5 N -2\over 48}$ & $\frac{1}{12} (N-1) (2 N+1)$ &No    \\ \hline

$C_{N}$(N~even)& ${N\over2}+1$&$(N,N-2, \ldots, 2,{N+2\over2})$ & ${7 N^2+ 19 N +10 \over48}$ & $\frac{1}{12} (N+2) (2 N+1)$ &No   \\ \hline
$C_{N}$(N~odd)& ${N\over2}+1$&$(N,N-2, \ldots, 3,{N+2\over2})$ & ${7 N^2+19 N+2\over 48}$ & $\frac{1}{12} \left(2 N^2+5 N+1\right)$ &Yes   \\ \hline

$D_{N}$(N~even)& $N-2$&$(N-2, N-4,\ldots, 2)$ & ${ 7 N^2-19 N+10 \over 48}$ & $\frac{1}{12} (N-2) (2 N-1)$ &No   \\ \hline
$D_{N}$(N~odd)& $N-2$&$({N-2},N-4, \ldots, 3)$ & ${ 7 N^2-19 N+2\over 48}$ & $\frac{1}{12} \left(2 N^2-5 N+1\right)$ &Yes    \\ \hline

$E_{6}$& $3$&$(3)$ & ${41\over 24}$ & ${13\over 6}$ &No    \\ \hline

$E_{7}$& $4$&$(4)$ & ${59\over24}$ & ${19\over 6}$ &No    \\ \hline

$E_{8}$& $6$&$(6)$ & ${95\over24}$ & ${31\over 6}$ &No    \\ \hline

$G_{2}$& $2$&$(2)$ & ${23\over 24}$ & ${7\over6}$ &No    \\ \hline

$F_{4}$& $3$&$(3)$ & ${41\over 24}$ & ${13\over 6}$ &No    \\ \hline

\end{tabular}
\end{center}
\caption{Physical data for $\mathcal{N}=2$ SCFT with minimal flavor central charge among theories constructed using $(2,0)$ construction.}
\label{data}
\end{table}
The flavor central charge can be put in the following form:
\begin{align}
& k_G=h^{\vee}-{1\over n}h,~~~~~~G=B_N,F_2,G_2,  \nonumber\\
& k_G=h^{\vee}-{1\over 2n}h,~~~~~G=C_N, \nonumber\\
&k_G={h\over 2},~~~~~~~~~~~~~~~~G=A_{N-1}, \nonumber\\
&k_G={h-2\over 2},~~~~~~~~~~~G=D_N, \nonumber\\
& k_G={h\over 6}+1,~~~~~~~~~~~G=E_N.
\label{central}
\end{align}
Here $n$ is the lacety of the Lie algebra $G$, see table. \ref{lie} for these numbers, $h^{\vee}$  the dual Coxeter number, and $h$ the Coxeter number.
For the theory without extra $U(1)$ flavor symmetry, the corresponding 2d chiral algebra \cite{Beem:2013sza} is given by the Kac-Moody algebra of type $G$ with level
\begin{equation}
k_{2d}=-k_G.
\end{equation}
The corresponding level is admissible only for $ A_{N-1} (N~odd)$ cases \cite{arakawa2015associated}.

The Higgs branch of these theories can be found from the associated variety of the corresponding vertex operator algebra \cite{Song:2017oew, Beem:2017ooy}.
For the theory without extra $U(1)$ flavor symmetry and if the level is admissible, the associated variety is found in \cite{arakawa2015associated}, and 
they are given by the nilpotent orbit  of the corresponding Lie algebra.  The associated variety of $E_6, E_7, E_8$ 
theory is also found in \cite{arakawa2016joseph}. For other cases, we use a simpler method by computing the Higgs branch dimension from the central charge data, and guess the 
corresponding orbit. See table. \ref{higgs} for the summary.
The Higgs branch is the minimal nilpotent orbit only for $ A_1, ,A_2, D_4, E_6, E_7,E_8$ case. 
\begin{table}[h]
\begin{center}
\begin{tabular}{|c|c|}
\hline
Flavor group & Nilpotent orbit  \\ \hline
$A_{N-1}$ ($N=2k+1$)& $[\underbrace{2,2,\ldots, 2}_{k},1]$   \\ \hline
$B_N$ (N odd)& $[3,\underbrace{2,\ldots, 2}_{N-3},1,1,1,1] $\\ \hline
$C_N$ (N even) & $[\underbrace{2,\ldots, 2}_{N-1},1,1]$ \\ \hline
$D_N$  (N even)& $[\underbrace{2,\ldots, 2}_{N-2},1,1,1,1]$\\ \hline
$E_N$ & $A_1$ \\ \hline
$G_2$ &$G_2(a_1)$ \\ \hline
$F_4$ & $\tilde{A}_1$ \\ \hline
\end{tabular}
\end{center}
\caption{The niplotent orbit for the Higgs branch of our minimal theory, and we list the partition for the classical group case, and the Bala-Carter label for exceptional group. }
\label{higgs}
\end{table}

This paper is organized as follows: section II gives the detailed construction for each minimal theory, we also study many interesting properties of these theories including 
the Coulomb branch spectrum, 2d chiral algebra, Higgs branch, etc; section III propose a conjecture about the lower bound of flavor central charge when the 
Coulomb branch spectrum has a common denominator $r$; Finally a conclusion is given in section IV.

\section{Minimal theory}
\subsection{ADE flavor group}
We can construct a large class of $\mathcal{N}=2$ SCFT by compactifying 6d $(2,0)$ theory of type $J=ADE$ on a sphere with one irregular singularity and one regular singularity  \footnote{The flavor central charge for theory defined using only regular singularity is simple: $k_G=h^{\vee}$ with $h^{\vee}$ the dual Coxeter number \cite{Gaiotto:2009we,Chacaltana:2012zy}. }. 
We are interested in following irregular singularity:
\begin{equation}
\Phi= {T\over z^{2+k/b}}+\ldots
\end{equation}
Here $k> -b$ and $k,b$ is copime, and $T$ is regular semi-simple \footnote{It is also possible to have non-abelian flavor symmetry from irregular singularity \cite{Xie:2017aqx}, but the flavor central charge is bigger than 
dual Coxeter number.}. The allowed value of $b$ is \cite{Wang:2015mra}:
\begin{equation}
A_{N-1}:~b|[N,N-1],~~D_{N}:~b|[2N-2, N],~~E_6:~~b|[12,9,8],~E_7:~b|[18,14],~E_8:~b|[30,24,20].
\end{equation}
Here the notation means that $b$ is the divisor of the listed integers in square bracket. The SW curve is found from the spectral curve of Hitchin system: $det(x-\Phi(z))=0$. They can be put in following form:
\begin{align}
& J=A_{N-1}:~~x^N+\sum_{i=2}^N \phi_i(z) x^{N-i}=0, \nonumber\\
& J=D_{N}:~~x^{2N}+\sum_{i=1}^{N-1} x^{2N-2i}+(\tilde{\phi}_N)^2=0, \nonumber\\
& J=E_6:~~\phi_2(z), \phi_5(z), \phi_6(z), \phi_8(z), \phi_9(z), \phi_{12}(z),         \nonumber\\
& J=E_7:~~ \phi_2(z), \phi_6(z), \phi_8(z), \phi_{10}(z), \phi_{12}(z), \phi_{14}(z), \phi_{18}(z),          \nonumber\\
& J=E_8:~~  \phi_2(z), \phi_8(z), \phi_{12}(z), \phi_{14}(z), \phi_{18}(z), \phi_{20}(z),   \phi_{24}(z),  \phi_{30}(z).            
\end{align}
Here $\phi_i(z)$ is a degree $i$ differential on Riemann surface. For $E_N$ case, we only list the independent differentials. The coefficients of these differentials encode the 
Coulomb branch spectrum of the theory.

We further assume that the regular puncture is regular semi-simple so that we have a ADE flavor symmetry. The flavor central charge of the theory is \cite{Xie:2013jc,Xie:2016evu,Xie:2017vaf,Xie:2017aqx}:
\begin{equation}
k_G=h-{b\over b+k}.
\end{equation}

We find the following minimal value of flavor central charge: 

$\bullet$ $G=A_{N-1}$: $b=N,b+k=2$. The flavor central charge is $k_G={N\over 2}$, and the Coulomb branch spectrum is $({N\over 2}, {N-2\over 2},\ldots, {3\over 2})$ for $N$ odd, and $({N\over 2}, {N-2\over 2},\ldots,2)$ for $N$ even. If N is even, there is a further $U(1)$ flavor symmetry and this theory is $SU({N\over 2})$ gauge theory
coupled with N flavor fundamental matter. 

$\bullet$ $G=D_N$: $b=N,b+k=1$.  ($N\geq 4$): The flavor central charge is $k_G=N-2$. The Coulomb branch spectrum is $(N-2,\ldots, 2)$ (N is even), or $(N-2,\ldots, 3)$ (N is odd). When N is even, this theory is $Usp(N-2)$ gauge theory coupled with $2N$ half fundamental flavors. When N is odd, there is 
an extra $U(1)$ flavor symmetry. 

$\bullet$ $G=E_6$: $b=9,b+k=1$. $k_G=3$. This is the rank one theory with Coulomb branch spectrum (3). The Higgs branch is the minimal nilpotent orbit of $E_6$.

$\bullet$ $G=E_7$: $b=14,b+k=1$. $k_G=4$. This is the rank one theory with Coulomb branch spectrum (4). The Higgs branch is the minimal nilpotent orbit of $E_7$.

$\bullet$ $G=E_8$: $b=24,b+k=1$. $k_G=6$. This is the rank one theory with Coulomb branch spectrum (6). The Higgs branch is the minimal nilpotent orbit of $E_8$.

$E_N$ type theory is the Minahan-Nemchesky theory \cite{Minahan:1996cj}.

\subsection{Non-simply laced flavor group}
To get non-simply laced flavor group from $(2,0)$ type construction, we need to do outer automorphism twist. The corresponding twist is listed in table. \ref{table:outerautomorphisms}:
\begin{table}[htb]
\begin{center}
  \begin{tabular}{ |c|c| c|c|c|c| }
    \hline
    $ J $ ~&$A_{2N}$ &$A_{2N-1}$ & $D_{N}$  &$E_6$&$D_4$ \\ \hline
   Outer Automorphism  &$Z_2$ &$Z_2$& $Z_2$  & $Z_2$&$Z_3$\\     \hline
        Invariant subalgebra  $g^\vee$ &$B_N$&$C_N$& $B_{N-1}$  & $F_4$&$G_2$\\     \hline
  \end{tabular}
\end{center}
\caption{Outer-automorphisms of simple Lie algebras and its invariant subalgebra.}
\label{table:outerautomorphisms}
\end{table}

$\bullet$ $G=B_{N}$: We use $Z_2$ twist of $A_{2N-1}$ theory. The action on the differential is $\phi_k\rightarrow (-1)^k \phi_k$ \cite{Chacaltana:2012ch}. This means that even degree differentials are holomorphic polynomial, and odd degree
differentials have half-integral order pole. Therefore, we can only have one class of theory (unlike the untwisted theory),
and the flavor central charge is given by the following formula
\begin{equation}
class~I:~k_G=2N-1-{1\over2}{2N\over 2N+k}.
\end{equation} 
The minimal theory is achieved at $k+2N=1$, and $k_G=N-1$. The Coulomb branch spectrum is then $(N-1,N-3,\ldots, 2)$ for N odd. For N even, we have the spectrum $(N-1, N-3,\ldots,3)$, and there is an extra
$U(1)$ flavor symmetry.

$\bullet$ $G=C_{N}$: We use $Z_2$ twist of $D_{N+1}$ theory. The non-trivial action on differential is $\tilde{\phi}_N\rightarrow -\tilde{\phi}_N$ \cite{Chacaltana:2013oka}. There are two class of theories, and the flavor central charge is given by following formula:
\begin{equation}
class~I:~k_G={1\over2}(2N+2-{2N\over 2N+k}),~class~II~:k_G={1\over2}(2N+2-{2N+2\over 2N+2k+3}).
\end{equation}  
The minimal theory is achieved for class I theory with $k+2N=2$, and $k_G={1\over2}N+1$. The Coulomb branch spectrum is $(N, N-2,\ldots, 2, {N+2\over 2})$ if N is even, which is just
a $SO(N+2)$ gauge theory with $2N$ half fundamental hypermultiplet. The Coulomb branch spectrum is $(N,N-2,\ldots, 3, {N+2\over 2})$ if $N$ is odd, but there is an extra $U(1)$ flavor symmetry for this theory.

$\bullet$ $G=G_2$: We use $Z_3$ twist of $D_{4}$ theory \cite{Chacaltana:2016shw}. The basis of differential is $(\phi_2, \phi_4^{\omega},\phi_4^{\omega^2},\phi_6^{'})$, and $\phi_4^{\omega}, \phi_4^{\omega^2}$ transform nontrivially under $Z_3$ group, and their 
order of pole has the form $z^{n+j/3}$. We only have one class of theory with the flavor central charge
\begin{equation}
k_G=4-{1\over3}{6\over k+6}.
\end{equation}
The minimal theory is achieved at $k+6=1$. The minimal theory has Coulomb branch spectrum  $2$. 

$\bullet$ $G=F_4$: We use $Z_2$ twist of $E_6$ theory \cite{Chacaltana:2015bna}. The $Z_2$ action on the differential is $\phi_5\rightarrow -\phi_5$ and $\phi_9\rightarrow - \phi_9$, and so they have half-integer order of pole. 
There are two classes of theories with following flavor central charge:
\begin{equation}
class~I:~k_G=9-{1\over2}{12\over k+12},~class~II~:k_G=9-{1\over2}{8\over k+8}.
\end{equation} 
The minimal theory is achieved in class I by taking $k+12=1$. The Coulomb branch  spectrum  is $3$.

\subsection{2d chiral algebra, central charges and Higgs branch}
There is a correspondence between Schur sector of 4d $\mathcal{N}=2$ SCFT and 2d vertex operator algebra \cite{Beem:2013sza}. For our 
theory, the corresponding 2d vertex operator algebra is just the Kac-Moody algebra of type $G$ \cite{Xie:2016evu,Song:2017oew,xie:2017dd}, and the level of 2d theory is 
\begin{equation}
k_{2d}=-k_G.
\end{equation}
See formula (\ref{central}) for flavor central charge of 4d theory. 
If our theory has an extra abelian flavor symmetry, then we need to add a $U(1)$ Kac-Moody algebra too.  A level is called admissible if 
\begin{align}
& k=-h+{p\over q},~~(p,q)=1,~~p\geq h, ~~G=ADE, \nonumber \\
& k=-h^{\vee}+{p\over q},~~(p,q)=1,~~p\geq h^{\vee},~~G=BCFG,  \nonumber\\
& k=-h^{\vee}+{p\over n q},~~(p,q)=1,~~(p,n)=1,~~p\geq h,~~G=BCFG.  
\end{align}
Here $h$ is the Coxeter number, $h^{\vee}$ is the dual Coxeter number, $n$ is the lacety of the Lie algebra, see table. \ref{lie}. 
The Higgs branch is the associated variety of the corresponding Kac-Moody algebra of type $G$, and the associated variety for admissible level is found in \cite{arakawa2015associated}.
In our case, only the level corresponding to $A_N(N~odd)$ case is admissible, and one can use the result of \cite{arakawa2015associated} to find the corresponding Higgs 
branch which is listed in table. \ref{higgs}. The corresponding associated vaiety for $E_N$ type is also considered in \cite{arakawa2016joseph}.

Using above 2d/4d relation, we propose the following central charge formula \cite{xie:2017dd}:
\begin{equation}
c= {1\over12}{-k_G dim(G) \over -k_G+h^{\vee}}-{f\over 12},~~2a-c={1\over 4}(\sum 2[u_i]-1). 
\end{equation}
here $h^{\vee}$ is the dual Coxeter number, and $f$ is the number of abelian flavor symmetry, and the second formula is found in \cite{Shapere:2008zf}. The corresponding Lie group data 
is listed in table. \ref{lie}. Using above formula, we find the central charges of our theory which are listed in table. \ref{data}.  Once we find the central charge $(a,c)$, we use the following formula
\begin{equation}
(a-c)=-{dim(Higgs)\over 24}
\end{equation}
to find out the dimension of Higgs branch and then the corresponding nilpotent orbit, which is listed in table. \ref{higgs}.

\begin{table}[h]
\begin{center}
\begin{tabular}{|c|c|c|c|c|}
\hline
~&dimension & $h$ & $h^{\vee}$&$n$ \\ \hline
$A_{N-1}$&$N^2-1$& $N$ & $N$&1  \\ \hline
$B_N$ & $(2N+1)N$ & $2N$& $2N-1$&2 \\ \hline
$C_N$ &  $(2N+1)N$&  $2N$ &. $N+1$&2 \\ \hline
 $D_N$ & $N(2N-1)$ & $2N-2$ & $2N-2$&1 \\ \hline
$E_6$& 78 & 12 & 12 &1\\ \hline
$E_7$& 133 & 18 & 18 &1\\ \hline
$E_8$& 248 & 30 & 30 &1\\ \hline
$F_4$& 52 & 12 & 9 &2\\ \hline
$G_2$& 14 & 6 & 4&3 \\ \hline

\end{tabular}
\end{center}
\caption{Lie algebra data, here $h$ is the Coexter number and $h^{\vee}$ is the dual Coexter number.}

\label{lie}
\end{table}

\subsection{Rank one theory}
Among our minimal theory, the rank one theory are listed in the following list
\begin{align}
& A_1,A_2, D_4,E_6,E_7,E_8, \nonumber\\
& B_3, B_4\times U(1), C_1\times U(1), G_2, F_4. 
\end{align}
We include $A_1$ case here (since among our minimal theory list, it is a free theory, so we have to search again, and it is easy to find theory with minimal flavor symmetry which is called $(A_1, A_3)$ theory or $H_1$ theory).
Here $B_3, G_2$ theroy have the same central charge $(a,c)$ as the $D_4$ theory, and $B_4\times U(1), F_4$ theory have the same $(a,c)$ values as $E_6$ theory, and $C_1\times U(1)$ have the same $(a,c)$ value as $A_2$ theory.

\section{A conjecture}
The flavor central charge of our theory takes the following general form:
\begin{align}
& k_G=h^{\vee}-{1\over n}{b\over k+b}.
\label{fla}
\end{align}
Here $n$ is the lacety of Lie group $G$. The allowed values  of $b$ \footnote{We do not require $(k,b)$ coprime here and consider only the case where the irregular singularity is regular semi-simple, and these restrictions do not lose any generality in considering the lower bound of flavor central charge.} are listed in table. \ref{bvalue}. 
\begin{table}[h]
\begin{center}
\begin{tabular}{|c|c|c|c|c|c|}
\hline
$A_{N-1}$&$b=[N,N-1]$ & $D_N$ & $b=[2N-2, N]$ &$E_6$ & $b=[12,9,8]$   \\ \hline
$E_7$ & $b=[18,14]$ &$E_8$ & $b=[30,24,20]$ &$B_N$ & $b=[2N]$ \\ \hline
$C_N$ & $b=[2N, 2N+2]$ & $G_2$ & $b=[6]$ & $F_4$ & $b=[12,8]$ \\ \hline
\end{tabular}
\caption{The allowed value of $b$ for flavor central charges appearing in formula \ref{fla}.}
\end{center}
\label{bvalue}
\end{table}
 
The common denominator of Coulomb branch operators in our theory is $r=k+b$. 
For $r\neq 1$, the minimal value is achieved for $b=h, k+b=r$. 
Our conjecture is that if the Coulomb branch has common denominator $r$, the flavor central charge has the following bound:
\begin{equation}
 k_G \geq h^{\vee}-{1\over n} {h\over r},
\end{equation}
for $r>1$. For $r=1$, we have 
\begin{equation}
k_G\geq k_{min},
\end{equation}
and $k_{min}$ is the value listed in table. \ref{data}.

\section{Conclusion}
We performed a scan of $\mathcal{N}=2$ SCFT with minimal flavor central charge  from theories constructed using 6d $(2,0)$ theory. Let's make 
some remarks about these theories:
\begin{itemize}
\item A first question is the uniqueness of theory with minimal flavor central charge. We conjecture that they are unique.

\item An interesting fact is that some of the theories discussed in this paper has to have extra abelian flavor symmetry. It is interesting to find out 
whether we can find the theory with only the simple flavor group. 

\item Our bound on flavor central charge is weaker than what is found in \cite{Beem:2013sza} for $G=BFG$ case. 

\item We find rank one theory with $B_3,G_2,F_4$ flavor group. Similar theory is also proposed in \cite{Argyres:2016yzz}. The central charges $(a,c)$ agree, but 
the Coulomb branch spectrum is different (their value is twice of ours). It would be interesting to find out whether these two sets of theories are the same or not.
We also find new rank one theory with $B_4\times U(1)$ and $C_1\times U(1)$ flavor symmetries (their flavor symmetry might be further enhanced.).

\item A quite peculiar feature for $B_3$ and $G_2$ theory is that they have a dimension two operator and therefore an exact marginal deformation, 
but there is no obvious weakly coupled gauge theory description. A relevant feature is that the exact marginal deformation appears in the differential which 
transforms non-trivially under the outer automorphism twist. 

\end{itemize}

\section*{Acknowledgements}
We would like to thank Yongchao Lu, Yifan Wang and Ke Ye for helpful discussions. 
The work of DX is supported by Center for Mathematical Sciences and Applications at Harvard University, and in part by the Fundamental Laws Initiative of
the Center for the Fundamental Laws of Nature, Harvard University.

\bibliographystyle{JHEP}
\bibliography{ref}

\providecommand{\href}[2]{#2}\begingroup\raggedright\begin{thebibliography}{10}

\bibitem{Hofman:2008ar}
D.~M. Hofman and J.~Maldacena, {\it {Conformal collider physics: Energy and
  charge correlations}},  {\em JHEP} {\bf 05} (2008) 012,
  [\href{http://xxx.lanl.gov/abs/0803.1467}{{\tt arXiv:0803.1467}}].

\bibitem{Liendo:2015ofa}
P.~Liendo, I.~Ramirez, and J.~Seo, {\it {Stress-tensor OPE in $ \mathcal{N}=2 $
  superconformal theories}},  {\em JHEP} {\bf 02} (2016) 019,
  [\href{http://xxx.lanl.gov/abs/1509.0003}{{\tt arXiv:1509.0003}}].

\bibitem{Argyres:1995jj}
P.~C. Argyres and M.~R. Douglas, {\it {New phenomena in SU(3) supersymmetric
  gauge theory}},  {\em Nucl. Phys.} {\bf B448} (1995) 93--126,
  [\href{http://xxx.lanl.gov/abs/hep-th/9505062}{{\tt hep-th/9505062}}].

\bibitem{Beem:2013sza}
C.~Beem, M.~Lemos, P.~Liendo, W.~Peelaers, L.~Rastelli, and B.~C. van Rees,
  {\it {Infinite Chiral Symmetry in Four Dimensions}},  {\em Commun. Math.
  Phys.} {\bf 336} (2015), no.~3 1359--1433,
  [\href{http://xxx.lanl.gov/abs/1312.5344}{{\tt arXiv:1312.5344}}].

\bibitem{Lemos:2015orc}
M.~Lemos and P.~Liendo, {\it {$\mathcal{N}=2$ central charge bounds from $2d$
  chiral algebras}},  {\em JHEP} {\bf 04} (2016) 004,
  [\href{http://xxx.lanl.gov/abs/1511.0744}{{\tt arXiv:1511.0744}}].

\bibitem{Argyres:1995xn}
P.~C. Argyres, M.~R. Plesser, N.~Seiberg, and E.~Witten, {\it {New N=2
  superconformal field theories in four-dimensions}},  {\em Nucl. Phys.} {\bf
  B461} (1996) 71--84, [\href{http://xxx.lanl.gov/abs/hep-th/9511154}{{\tt
  hep-th/9511154}}].

\bibitem{Minahan:1996cj}
J.~A. Minahan and D.~Nemeschansky, {\it {Superconformal fixed points with E(n)
  global symmetry}},  {\em Nucl. Phys.} {\bf B489} (1997) 24--46,
  [\href{http://xxx.lanl.gov/abs/hep-th/9610076}{{\tt hep-th/9610076}}].

\bibitem{Xie:2012hs}
D.~Xie, {\it {General Argyres-Douglas Theory}},  {\em JHEP} {\bf 01} (2013)
  100, [\href{http://xxx.lanl.gov/abs/1204.2270}{{\tt arXiv:1204.2270}}].

\bibitem{Wang:2015mra}
Y.~Wang and D.~Xie, {\it {Classification of Argyres-Douglas theories from M5
  branes}},  {\em Phys. Rev.} {\bf D94} (2016), no.~6 065012,
  [\href{http://xxx.lanl.gov/abs/1509.0084}{{\tt arXiv:1509.0084}}].

\bibitem{arakawa2015associated}
T.~Arakawa, {\it Associated varieties of modules over kac--moody algebras and c
  2-cofiniteness of w-algebras},  {\em International Mathematics Research
  Notices} {\bf 2015} (2015), no.~22 11605--11666.

\bibitem{Song:2017oew}
J.~Song, D.~Xie, and W.~Yan, {\it {Vertex operator algebras of Argyres-Douglas
  theories from M5-branes}},  \href{http://xxx.lanl.gov/abs/1706.0160}{{\tt
  arXiv:1706.0160}}.

\bibitem{Beem:2017ooy}
C.~Beem and L.~Rastelli, {\it {Vertex operator algebras, Higgs branches, and
  modular differential equations}},
  \href{http://xxx.lanl.gov/abs/1707.0767}{{\tt arXiv:1707.0767}}.

\bibitem{arakawa2016joseph}
T.~Arakawa and A.~Moreau, {\it Joseph ideals and lisse minimal $ w $-algebras},
   {\em Journal of the Institute of Mathematics of Jussieu} (2016) 1--21.

\bibitem{Gaiotto:2009we}
D.~Gaiotto, {\it {N=2 dualities}},  {\em JHEP} {\bf 08} (2012) 034,
  [\href{http://xxx.lanl.gov/abs/0904.2715}{{\tt arXiv:0904.2715}}].

\bibitem{Chacaltana:2012zy}
O.~Chacaltana, J.~Distler, and Y.~Tachikawa, {\it {Nilpotent orbits and
  codimension-two defects of 6d N=(2,0) theories}},  {\em Int. J. Mod. Phys.}
  {\bf A28} (2013) 1340006, [\href{http://xxx.lanl.gov/abs/1203.2930}{{\tt
  arXiv:1203.2930}}].

\bibitem{Xie:2017aqx}
D.~Xie and K.~Ye, {\it {Argyres-Douglas matter and S-duality: Part II}},
  \href{http://xxx.lanl.gov/abs/1711.0668}{{\tt arXiv:1711.0668}}.

\bibitem{Xie:2013jc}
D.~Xie and P.~Zhao, {\it {Central charges and RG flow of strongly-coupled N=2
  theory}},  {\em JHEP} {\bf 03} (2013) 006,
  [\href{http://xxx.lanl.gov/abs/1301.0210}{{\tt arXiv:1301.0210}}].

\bibitem{Xie:2016evu}
D.~Xie, W.~Yan, and S.-T. Yau, {\it {Chiral algebra of Argyres-Douglas theory
  from M5 brane}},  \href{http://xxx.lanl.gov/abs/1604.0215}{{\tt
  arXiv:1604.0215}}.

\bibitem{Xie:2017vaf}
D.~Xie and S.-T. Yau, {\it {Argyres-Douglas matter and N=2 dualities}},
  \href{http://xxx.lanl.gov/abs/1701.0112}{{\tt arXiv:1701.0112}}.

\bibitem{Chacaltana:2012ch}
O.~Chacaltana, J.~Distler, and Y.~Tachikawa, {\it {Gaiotto duality for the
  twisted A$_{2N?1}$ series}},  {\em JHEP} {\bf 05} (2015) 075,
  [\href{http://xxx.lanl.gov/abs/1212.3952}{{\tt arXiv:1212.3952}}].

\bibitem{Chacaltana:2013oka}
O.~Chacaltana, J.~Distler, and A.~Trimm, {\it {Tinkertoys for the Twisted
  D-Series}},  \href{http://xxx.lanl.gov/abs/1309.2299}{{\tt arXiv:1309.2299}}.

\bibitem{Chacaltana:2016shw}
O.~Chacaltana, J.~Distler, and A.~Trimm, {\it {Tinkertoys for the Z3-twisted D4
  Theory}},  \href{http://xxx.lanl.gov/abs/1601.0207}{{\tt arXiv:1601.0207}}.

\bibitem{Chacaltana:2015bna}
O.~Chacaltana, J.~Distler, and A.~Trimm, {\it {Tinkertoys for the Twisted $E_6$
  Theory}},  {\em JHEP} {\bf 04} (2015) 173,
  [\href{http://xxx.lanl.gov/abs/1501.0035}{{\tt arXiv:1501.0035}}].

\bibitem{xie:2017dd}
Y.~Wang and D.~Xie, {\it {To appear}}, .

\bibitem{Shapere:2008zf}
A.~D. Shapere and Y.~Tachikawa, {\it {Central charges of N=2 superconformal
  field theories in four dimensions}},  {\em JHEP} {\bf 09} (2008) 109,
  [\href{http://xxx.lanl.gov/abs/0804.1957}{{\tt arXiv:0804.1957}}].

\bibitem{Argyres:2016yzz}
P.~C. Argyres and M.~Martone, {\it {4d $ \mathcal{N} $ =2 theories with
  disconnected gauge groups}},  {\em JHEP} {\bf 03} (2017) 145,
  [\href{http://xxx.lanl.gov/abs/1611.0860}{{\tt arXiv:1611.0860}}].

\end{thebibliography}\endgroup

\end{document}